\documentclass[prd,aps,twocolumn,showpacs,nofootinbib]{revtex4-1}

\usepackage{amsmath,amssymb,amsthm,wasysym}
\usepackage{graphicx}
\usepackage{psfrag}
\usepackage{epstopdf}
\usepackage{color}
\usepackage{mathrsfs}
\usepackage{fixmath}

\DeclareSymbolFontAlphabet{\mathrsfs}{rsfs}
\DeclareMathAlphabet{\mathcal}{OMS}{cmsy}{m}{n}

\allowdisplaybreaks
\newcommand{\ah}{{\hat{a}}}
\newcommand{\ahstar}{{\hat{a}^*}}

\newcommand{\be}{\begin{equation}}
\newcommand{\bl}{\mbox{\boldmath$\ell$}}
\newcommand{\bse}{\begin{subequations}}
\newcommand{\bsigma}{\mbox{\boldmath$\sigma$}}

\newcommand{\cpipii}{{\mathbf{p}}_1\times{\mathbf{p}}_2}

\newcommand{\ee}{\end{equation}}
\newcommand{\ese}{\end{subequations}}

\newcommand{\ncpi}{{\mathbf{n}}_{12}\times{\mathbf{p}}_1}
\newcommand{\ncpii}{{\mathbf{n}}_{12}\times{\mathbf{p}}_2}
\newcommand{\np}{({\bf n}\cdot{\bf p})}
\newcommand{\npi}{(\mathbf{n}_{12}\cdot\mathbf{p}_1)}
\newcommand{\npii}{({\bf n}_{12}\cdot{\bf p}_2)}

\newcommand{\piipii}{{\bf p}_2^2}

\newcommand{\pipi}{{\bf p}_1^2}
\newcommand{\pipii}{({\mathbf{p}}_1\cdot{\mathbf{p}}_2)}

\newcommand{\pp}{{\bf p}^2}

\def\vct#1{\mathbf{#1}}

\def\nunit{n}

\newcommand{\vnunit}{\vct{\nunit}_{12}}

\def\canmom{p}

\newcommand{\vmom}[1]{{\vct{\canmom}}_{#1}}

\newcommand{\vnun}{{\vnunit}}

\newcommand{\scpm}[2]{\left(#1\cdot#2\right)}
\newcommand{\vspin}[1]{\hat{\vct{S}}_{1}}

\def\gravthree{G}

\def\rel{r_{12}} 

\def\G{{\bar{G}}}

\def\de{\partial}
\def\r{{\mathbf r}}
\def\p{{\mathbf p}}
\def\x{{\mathbf x}}
\def\n{{\mathbf n}}
\def\S{{\bar{\mathbf S}}}
\def\Sstar{{\bar{\mathbf S}}^*}
\def\Omg{{\mathbf\Omega}}

\def\de{\partial}

\def\H{\hat{H}}

\def\V{{\bf V}}

\begin{document}

\title{Effective one body Hamiltonian of two spinning black-holes
with next-to-next-to-leading order spin-orbit coupling}

\author{Alessandro \surname{Nagar}}
\affiliation{Institut des Hautes Etudes Scientifiques, 91440
  Bures-sur-Yvette, France}

\date{\today}

\begin{abstract}
Building on the recently computed next-to-next-to-leading order (NNLO) post-Newtonian (PN) spin-orbit Hamiltonian 
for spinning binaries~\cite{Hartung:2011te} we improve the effective-one-body (EOB) description of the dynamics 
of two spinning black-holes  by including NNLO effects in the spin-orbit interaction. 
The calculation that is presented extends to NNLO the next-to-leading order (NLO) spin-orbit Hamiltonian 
computed in Ref.~\cite{Damour:2008qf}. 
The present EOB Hamiltonian reproduces the spin-orbit coupling  through NNLO in the test-particle limit case. 
In addition, in the case of spins parallel or antiparallel to the orbital angular momentum, when circular 
orbits exist, we find that the inclusion of NNLO spin-orbit terms moderates the effect of the NLO spin-orbit coupling.
\end{abstract}


\maketitle

\section{Introduction}
Coalescing black-hole binaries are among the most promising gravitational wave (GW) 
sources for the currently operating network of ground-based interferometric GW detectors.
Since the spin-orbit interaction can increase the binding energy of the last stable
orbit, and thereby leading to large GW emission, it is reasonable to think that the 
first detections will concern binary systems made of spinning binaries. For this reason,
there is a urgent need of template waveforms accurately describing the GW emission
from coalescing spinning black-hole binaries. These template waveforms will be functions
of at least eight intrinsic real parameters: the two masses $m_1$ and $m_2$ and the 
two spin-vectors ${\bf S}_1$ and ${\bf S}_2$. Because of the multidimensionality of 
the parameter space, it seems unlikely for state-of-the-art numerical simulations
to densely sample this parameter space. This gives a boost to develop
{\it analytical} methods for computing the needed, densely spaced, bank of 
accurate template waveforms.
Among the existing analytical methods for computing the motion and the dynamics of
black hole (and neutron star) binaries, the most complete and the most promising 
is the effective-one-body approach 
(EOB)~\cite{Buonanno:1998gg,Buonanno:2000ef,Damour:2000we,Damour:2001tu,Damour:2008gu,
Damour:2009wj}.
Several recent works have shown the possibility of getting an excellent agreement
between the EOB analytical waveforms and the outcome of numerical simulations
of coalescing black-hole (and inspiralling neutron-star~\cite{Baiotti:2010xh,Baiotti:2011am}) 
binaries. A considerable part of the current literature deals with nonspinning black-hole 
systems~\cite{Buonanno:2006ui,Buonanno:2007pf,Damour:2007yf,Damour:2007vq,Boyle:2008ge,Damour:2008te,
Damour:2009kr,Buonanno:2009qa}, with different (though not extreme) mass ratios 
(see in particular~\cite{Pan:2011gk,Tiec:2011bk}) or in the (circularized) extreme-mass-ratio 
limit~\cite{Nagar:2006xv,Damour:2007xr,Bernuzzi:2010ty,Bernuzzi:2010xj,Yunes:2009ef} 
(notably including spin~\cite{Yunes:2010zj}).

The work at the interface between numerical relativity and the analytical 
EOB description of spinning binaries has been developing fast in recent years. 
The first EOB Hamiltonian which included spin effects was conceived in 
Ref.~\cite{Damour:2001tu}. It was shown there that one could map the 3PN
dynamics, together with the leading-order (LO) spin-orbit and spin-spin
dynamical effects of a binary systems, onto an effective test-particle
moving in a Kerr-type metric, together with an additional spin-orbit
interaction. In Ref.~\cite{Buonanno:2005xu} the use of the nonspinning 
EOB Hamiltonian augmented with PN-type spin-orbit and spin-spin 
terms allowed to carry out the first (and up to now, only) analytical 
exploratory study of the dynamics and waveforms from coalescing spinning 
binaries with precessing spins. 
Recently, Ref.~\cite{Damour:2008qf}, building upon the PN-expanded Hamiltonian 
of~\cite{Damour:2007nc}, extended the EOB approach of~\cite{Damour:2001tu} 
so to include the next-to-leading-order (NLO) spin-orbit couplings
(see also Refs.~\cite{Faye:2006gx,Blanchet:2006gy} for a derivation of 
these couplings in the harmonic-coordinates equations of motion 
and Ref.~\cite{Levi:2010zu} for a derivation using an effective field theory approach). 
Using this model (with the addition of EOB-resummed radiation reaction 
force~\cite{Damour:2007xr,Damour:2008gu,Pan:2010hz}), Ref.~\cite{Pan:2009wj} 
performed the first comparison with numerical-relativity simulations of 
nonprecessing, spinning, equal-mass, black-holes binaries. Then,  
building on Ref.~\cite{Damour:2001tu,Damour:2008qf} and Ref.~\cite{Barausse:2009aa}, 
Ref.~\cite{Barausse:2009xi} worked out an improved Hamiltonian 
for spinning black-hole binaries. 

Recently, Hartung and Steinhoff~\cite{Hartung:2011te} have computed the PN-expanded 
spin-orbit Hamiltonian at next-to-next-to-leading order (NNLO), pushing one PN order 
further the previous computation of Damour, Jaranowski and Sch\"afer~\cite{Damour:2007nc}. 
The result of Ref.~\cite{Hartung:2011te} completes the knowledge of the PN
Hamiltonian for binary spinning black-holes up to and including 3.5PN.

This paper belongs to the lineage of Refs.~\cite{Damour:2001tu,Damour:2008qf} and
it aims at exploiting the PN-expanded Hamiltonian  of Ref.~\cite{Hartung:2011te}  
so as to obtain the NNLO-accurate spin-orbit interaction as it enters the EOB
formalism. Note that, by contrast to Refs.~\cite{Barausse:2009xi} and~\cite{Damour:2008qf}, 
we shall not discuss here spin-spin interactions, nor shall we try to propose 
a specific way to incorporate our NNLO spin-orbit results into some complete, 
resummed EOB Hamiltonian. 
Although the Hamiltonian that we shall discuss here does not resum all 
the spin-orbit terms entering the formal ``spinning test-particle limit'',
we shall check that it consistently reproduces the ``spinning test-particle'' 
results of Ref.~\cite{Barausse:2009xi}.

The paper is organized as follows: in Sec.~\ref{sec:PN} we recall the structure
of the PN-expanded spin-orbit Hamiltonian (in Arnowitt-Deser-Misner (ADM) coordinates) of 
Ref.~\cite{Hartung:2011te} and then we express it in the center of mass frame. 
Section~\ref{sec:Heff} explicitly performs the canonical transformation from
ADM coordinates to EOB coordinates and finally computes the effective 
Hamiltonian, and, in particular, the effective gyro-gravitomagnetic ratios.
In Sec.~\ref{sec:fixings} we discuss the case of circular equatorial orbits,
we derive the test-mass limit and we exploit the gauge freedom to simplify the
expression of the final Hamiltonian.

We adopt the notation of~\cite{Damour:2008qf} and we use the letters 
$a,b=1,2$ as particle labels. Then, $m_a$, $\mathbf{x}_a=(x_a^i)$, 
$\mathbf{p}_a=(p_{ai})$, and $\mathbf{S}_a=(S_{ai})$ denote, 
respectively, the mass, the position vector, the linear momentum 
vector, and the spin vector of the $a$th body;
for $a\ne b$ we also define
$\mathbf{r}_{ab}\equiv\mathbf{x}_a-\mathbf{x}_b$,
$r_{ab}\equiv|\mathbf{r}_{ab}|$,
$\mathbf{n}_{ab}\equiv\mathbf{r}_{ab}/r_{ab}$,
$|\cdot|$ stands here for the Euclidean length of 
a 3-vector.

\section{PN-expanded Hamiltonian in ADM coordinates}
\label{sec:PN}
We closely follow the procedure of Ref.~\cite{Damour:2008qf}. 
The starting point of the calculation is the PN-expandend two-body Hamiltonian 
$H$ which can be decomposed as the sum of an orbital part, $H_{\rm o}$, 
a spin-orbit part, $H_{\rm so}$ (linear in the spins) and a spin-spin term 
$H_{\rm ss}$ (quadratic in the spins), that we quote here for completeness
but that we are not going to discuss in the paper. It reads
\begin{align}
\label{fullH}
H(\mathbf{x}_a,\mathbf{p}_a,{\bf S}_a)
&= H_{\mathrm{o}}({\bf x}_a,{\bf p}_a)
+ H_{\mathrm{so}}({\bf x}_a,{\bf p}_a,{\bf S}_a)
\nonumber\\[1ex]&\quad
+ H_{\mathrm{ss}}({\bf x}_a,{\bf p}_a,{\bf S}_a).
\end{align}
The orbital Hamiltonian $H_{\mathrm{o}}$ includes the rest-mass
contribution and is explicitly known (in ADM-like coordinates)
up to the 3PN order~\cite{Damour:2000kk,Damour:2001bu}.
It has the structure
\begin{align}
\label{horb}
H_{\mathrm{o}}({\bf x}_a,{\bf p}_a) &= \sum_a m_a c^2
+ H_{\text{oN}}({\bf x}_a,{\bf p}_a)
\nonumber\\[1ex]&\quad
+ \frac{1}{c^2}\,H_{\rm o1PN}({\bf x}_a,{\bf p}_a)
+ \frac{1}{c^4}\,H_{\rm o2PN}({\bf x}_a,{\bf p}_a)
\nonumber\\[1ex]&\quad
+ \frac{1}{c^6}\,H_{\rm o3PN}({\bf x}_a,{\bf p}_a)
+ {\cal O}\left(\frac{1}{c^8}\right).
\end{align}
The spin-orbit Hamiltonian $H_{\rm so}$ can be written as
\be
H_{\rm so}(\x_a,\p_a,{\mathbf S}_a)=\sum_a \Omg_a(\x_b,\p_b)\cdot{\mathbf S}_a.
\ee
Here, the quantity $\Omg_a$ is the sum of three contributions: the LO ($\propto 1/c^2$), 
the NLO ($\propto 1/c^4$), and the NNLO one ($\propto 1/c^6$), 
\be
\Omg_a(\x_b,\p_b)=\Omg^{\rm LO}_a(\x_b,\p_b)+\Omg^{\rm NLO}_a(\x_b,\p_b)+\Omg^{\rm NNLO}_a(\x_b,\p_b).
\ee
The 3-vectors $\Omg_a^{\rm LO}$ and $\Omg_a^{\rm NLO}$ were explicitly computed 
in Ref.~\cite{Damour:2007nc}, while $\Omg_a^{\rm NNLO}$ can be read off Eq.(5) 
of Ref.~\cite{Hartung:2011te}. We write them here explicitly for completeness. 
For the particle label $a=1$, we have
\begin{widetext}
\begin{subequations}
\label{omegafinal}
\begin{align}
\mathbf{\Omega}^{\rm LO}_{1} &= \frac{G}{c^2r_{12}^2}
\bigg( \frac{3m_2}{2m_1}\ncpi - 2\ncpii \bigg),
\\[2ex]
\mathbf{\Omega}^{\rm NLO}_{1} &= \frac{G^2}{c^4r_{12}^3} \Bigg(
\bigg(-\frac{11}{2}m_2-5\frac{m_2^2}{m_1}\bigg)\ncpi
+ \bigg(6m_1+\frac{15}{2}m_2\bigg)\ncpii \Bigg)
\nonumber\\[1ex]&\quad
+ \frac{G}{c^4r_{12}^2} \Bigg( \bigg(
- \frac{5m_2\pipi}{8m_1^3} - \frac{3\pipii}{4m_1^2}
+ \frac{3\piipii}{4m_1m_2}
- \frac{3\npi\npii}{4m_1^2} - \frac{3\npii^2}{2m_1m_2} \bigg)\ncpi
\nonumber\\[1ex]&\quad
+ \bigg(\frac{\pipii}{m_1m_2}+\frac{3\npi\npii}{m_1m_2}\bigg)\ncpii
+ \bigg( \frac{3\npi}{4m_1^2} - \frac{2\npii}{m_1m_2} \bigg)\cpipii
\Bigg),
\\
\mathbf{\Omega}^{\rm NNLO}_1& = \frac{\gravthree}{\rel^2} \biggl[
	\biggl(
		\frac{7 m_2 (\vmom{1}^2)^2}{16 m_1^5} 
		+ \frac{9 \scpm{\vnun}{\vmom{1}}\scpm{\vnun}{\vmom{2}}\vmom{1}^2}{16 m_1^4} 
		+ \frac{3 \vmom{1}^2 \scpm{\vnun}{\vmom{2}}^2}{4 m_1^3 m_2}\nonumber\\
&		+ \frac{45 \scpm{\vnun}{\vmom{1}}\scpm{\vnun}{\vmom{2}}^3}{16 m_1^2 m_2^2}
		+ \frac{9 \vmom{1}^2 \scpm{\vmom{1}}{\vmom{2}}}{16 m_1^4}
		- \frac{3 \scpm{\vnun}{\vmom{2}}^2 \scpm{\vmom{1}}{\vmom{2}}}{16 m_1^2 m_2^2}\nonumber\\
&		- \frac{3 (\vmom{1}^2) (\vmom{2}^2)}{16 m_1^3 m_2}
		- \frac{15 \scpm{\vnun}{\vmom{1}}\scpm{\vnun}{\vmom{2}} \vmom{2}^2}{16 m_1^2 m_2^2}
		+ \frac{3 \scpm{\vnun}{\vmom{2}}^2 \vmom{2}^2}{4 m_1 m_2^3}\nonumber\\
&		- \frac{3 \scpm{\vmom{1}}{\vmom{2}} \vmom{2}^2}{16 m_1^2 m_2^2}
		- \frac{3 (\vmom{2}^2)^2}{16 m_1 m_2^3}
	\biggr)\vnun \times \vmom{1}
	+\biggl(
		- \frac{3 \scpm{\vnun}{\vmom{1}}\scpm{\vnun}{\vmom{2}}\vmom{1}^2}{2 m_1^3 m_2}\nonumber\\
&		- \frac{15 \scpm{\vnun}{\vmom{1}}^2\scpm{\vnun}{\vmom{2}}^2}{4 m_1^2 m_2^2}
		+ \frac{3 \vmom{1}^2 \scpm{\vnun}{\vmom{2}}^2}{4 m_1^2 m_2^2}
		- \frac{\vmom{1}^2 \scpm{\vmom{1}}{\vmom{2}}}{2 m_1^3 m_2}
		+ \frac{\scpm{\vmom{1}}{\vmom{2}}^2}{2 m_1^2 m_2^2}\nonumber\\
&		+ \frac{3 \scpm{\vnun}{\vmom{1}}^2 \vmom{2}^2}{4 m_1^2 m_2^2}
		- \frac{(\vmom{1}^2) (\vmom{2}^2)}{4 m_1^2 m_2^2}
		- \frac{3 \scpm{\vnun}{\vmom{1}}\scpm{\vnun}{\vmom{2}}\vmom{2}^2}{2 m_1 m_2^3}
		- \frac{\scpm{\vmom{1}}{\vmom{2}} \vmom{2}^2}{2 m_1 m_2^3}
	\biggr)\vnun \times \vmom{2}\nonumber\\
&	+\biggl(
		- \frac{9 \scpm{\vnun}{\vmom{1}} \vmom{1}^2}{16 m_1^4}
		+ \frac{\vmom{1}^2 \scpm{\vnun}{\vmom{2}}}{m_1^3 m_2}\nonumber\\
&		+ \frac{27 \scpm{\vnun}{\vmom{1}}\scpm{\vnun}{\vmom{2}}^2}{16 m_1^2 m_2^2}
		- \frac{\scpm{\vnun}{\vmom{2}}\scpm{\vmom{1}}{\vmom{2}}}{8 m_1^2 m_2^2}
		- \frac{15 \scpm{\vnun}{\vmom{1}} \vmom{2}^2}{16 m_1^2 m_2^2}\nonumber\\
&		+ \frac{\scpm{\vnun}{\vmom{2}}\vmom{2}^2}{m_1 m_2^3}
	\biggr)\vmom{1} \times \vmom{2}
\biggr] \nonumber\\
&+ \frac{\gravthree^2}{\rel^3} \biggl[
	\biggl(
		-\frac{3 m_2 \scpm{\vnun}{\vmom{1}}^2}{2 m_1^2}
		+\left(
			-\frac{3 m_2}{2 m_1^2}
			+\frac{27 m_2^2}{8 m_1^3}
		\right) \vmom{1}^2
		+\left(
			\frac{177}{16 m_1}
			+\frac{11}{m_2}
		\right) \scpm{\vnun}{\vmom{2}}^2\nonumber\\
&		+\left(
			\frac{11}{2 m_1}
			+\frac{9 m_2}{2 m_1^2}
		\right) \scpm{\vnun}{\vmom{1}} \scpm{\vnun}{\vmom{2}}
		+\left(
			\frac{23}{4 m_1}
			+\frac{9 m_2}{2 m_1^2}
		\right) \scpm{\vmom{1}}{\vmom{2}}\nonumber\\
&		-\left(
			\frac{159}{16 m_1}
			+\frac{37}{8 m_2}
		\right) \vmom{2}^2
	\biggr)\vnun \times \vmom{1}
	+\biggl(
		\frac{4 \scpm{\vnun}{\vmom{1}}^2}{m_1}
		+\frac{13 \vmom{1}^2}{2 m_1}\nonumber\\
&		+\frac{5 \scpm{\vnun}{\vmom{2}}^2}{m_2}
		+\frac{53 \vmom{2}^2}{8 m_2}
		- \left(
			\frac{211}{8 m_1}
			+\frac{22}{m_2}
		\right) \scpm{\vnun}{\vmom{1}} \scpm{\vnun}{\vmom{2}}\nonumber\\
&		-\left(
			\frac{47}{8 m_1}
			+\frac{5}{m_2}
		\right)\scpm{\vmom{1}}{\vmom{2}}
	\biggr)\vnun \times \vmom{2}\nonumber\\
&	+\biggl(
		-\left(
			\frac{8}{m_1}
			+\frac{9 m_2}{2 m_1^2}
		\right)\scpm{\vnun}{\vmom{1}}
		+\left(
			\frac{59}{4 m_1}
			+\frac{27}{2 m_2}
		\right)\scpm{\vnun}{\vmom{2}}
	\biggr)\vmom{1} \times \vmom{2}
\biggr]\nonumber\\
&+\frac{\gravthree^3}{\rel^4} \biggl[
	\left(
		\frac{181 m_1 m_2}{16}
		+ \frac{95 m_2^2}{4}
		+ \frac{75 m_2^3}{8 m_1}
	\right) \vnun \times \vmom{1}
	- \left(
		\frac{21 m_1^2}{2}
		+ \frac{473 m_1 m_2}{16}
		+ \frac{63 m_2^2}{4}
	\right)\vnun \times \vmom{2}
\biggr]
\end{align}
\end{subequations}
\end{widetext}
The expressions for $\mathbf{\Omega}^{\rm LO}_{2}$, $\mathbf{\Omega}^{\rm NLO}_{2}$ 
and $\mathbf{\Omega}^{\rm NNLO}_{2}$ can be obtained from the above formulas by
exchanging the particle labels 1~and~2.

Let us consider now the dynamics of the relative motion of the two body system
in the center of mass frame, which is defined by setting $\p_1+\p_2=0$.
Following~\cite{Damour:2008qf}, we rescale the phase-space variables 
${\bf R}\equiv {\bf x}_1-{\bf x}_2$ and ${\bf P}\equiv \p_1=-\p_2$
of the relative motion as
\begin{align}
\label{def1}
\r \equiv\dfrac{\bf R}{GM},\qquad
\p\equiv\dfrac{{\bf P}}{\mu}\equiv \dfrac{\p_1}{\mu}=-\dfrac{\p_2}{\mu},
\end{align}
where $M=m_1+m_2$ and $\mu\equiv m_1 m_2/M$. In addition, we rescale the original time
variable $T$ and any part of the Hamiltonian as 
\begin{equation}
\label{def2}
t\equiv \dfrac{T}{GM}, \quad \H^{\rm NR}\equiv \dfrac{H^{\rm NR}}{\mu,}
\end{equation}
where $H^{\rm NR}\equiv H-Mc^2$ denotes the ``nonrelativistic'' Hamiltonian, 
i.e. the Hamiltonian withouth the rest-mass contribution.
As in~\cite{Damour:2008qf} we work with the following two, basic combinations 
of the spin vectors:
\begin{align}
\label{def3}
{\bf S}   &\equiv{\bf S}_1 + {\bf S}_2 = m_1 c {\bf a}_1 + m_2 c {\bf a}_2,\\
\label{def4}
{\bf S}^* &\equiv \frac{m_2}{m_1}{\bf S}_1 + \frac{m_1}{m_2}{\bf S}_2  = m_2 c {\bf a}_1 + m_1 c {\bf a}_2,  
\end{align}
where we have also introduced the Kerr parameters of the individual black-holes,
$\mathbf{a}_1\equiv\mathbf{S}_1/(m_1c)$ and $\mathbf{a}_2\equiv\mathbf{S}_2/(m_2c)$.
We recall that in the formal~\footnote{As noted in Ref.~\cite{Damour:2008qf} this
formal limit is not relevant for the physically most important case of binary
black holes, for which ${\bf a}_2\to 0$ and $m_2\to 0$.} ``spinning test mass limit'' 
where, for example, $m_2\to0$ 
and ${\bf S}_2\to0$, while keeping ${\bf a}_2={\bf S}_2/(m_2c)$ fixed, 
one has a ``background mass'' $M \simeq m_1$, a ``background spin'' 
${\bf S}_{\rm bckgd}\equiv Mc\,{\bf a}_{\rm bckgd} \simeq{\bf S}_1=m_1c\,{\bf a}_1$,
a ``test mass'' $\mu \simeq m_2$, and a ``test spin''
$ {\bf S}_{\rm test}={\bf S}_2 =m_2c\,{\bf a}_2 \simeq \mu c\,{\bf a}_{\rm test}$
[with ${\bf a}_{\rm test}\equiv{\bf S}_{\rm test}/(\mu c)$].
Then, in this limit the combination
$\mathbf{S}\simeq\mathbf{S}_1=m_1c\,{\bf a}_1
\simeq Mc\,{\bf a}_{\rm bckgd}={\bf S}_{\rm bckgd}$
measures the background spin, while the other combination,
$\mathbf{S}^*\simeq m_1c\,{\bf a}_2\simeq Mc\,{\bf a}_{\rm test}
=M {\bf S}_{\rm test}/\mu$ measures the (specific) test spin
${\bf a}_{\rm test}={\bf S}_{\rm test}/ (\mu c)$.
Finally, since the use of the rescaled variables corresponds to a rescaling
of the action by a factor $1/(GM\mu)$, it is also natural to work
with the corresponding rescaled variables
\begin{equation}
\label{def5}
\S^{\rm X}\equiv \dfrac{{\bf S}^{\rm X}}{GM\mu},
\end{equation}
for any label X (X$=1,2,\;\; ,*$).

Using the definitions~\eqref{def1}-\eqref{def5}, the center-of-mass spin-orbit Hamiltonian
(divided by $\mu$) in terms of the rescaled variables has the structure
\begin{align}
\H_{\rm so}(\r,\p,\S,\Sstar) &\equiv  \dfrac{H_{\rm so}(\r,\p,\S,\Sstar)}{\mu}\\
                            &= \dfrac{1}{c^2}\H^{\rm so}_{\rm LO}(\r,\p,\S,\Sstar)\nonumber\\
                            &+ \dfrac{1}{c^4}\H^{\rm so}_{\rm NLO}(\r,\p,\S,\Sstar)\nonumber\\
                            &+ \dfrac{1}{c^6}\H^{\rm so}_{\rm NNLO}(\r,\p,\S,\Sstar)+{\cal O}\left(\dfrac{1}{c^8}\right),
\end{align}
and it can be written as
\be
\label{eq:Hso}
\H_{\rm so}(\r,\p,\S,\Sstar)=\dfrac{\nu}{c^2 r^2}\left(g_s^{\rm ADM}(\bar{S},n,p) + g_{S^*}^{\rm ADM}(\bar{S}^*,n,p)\right),
\ee
with the following definitions: $\nu\equiv\mu/M$ is the symmetric mass ratios 
and ranges from 0 (test-body limit) to 1/4 (equal-mass case); 
the notation $(V_1,V_2,V_3)\equiv \V_1\cdot (\V_2\times \V_3)=\epsilon_{ijk}V_1^iV_2^jV_3^k$ stands 
for the Euclidean mixed products of 3-vectors; $\n\equiv \r/|r|$; $g_S^{\rm ADM}$ and $g_{S^*}^{\rm ADM}$ 
are the two (dimensionless) gyro-gravitomagnetic ratios as introduced (up to NLO accuracy) in~\cite{Damour:2008qf}. 
These two coefficients parametrize the coupling between the spin vectors and the apparent gravito-magnetic 
field seen in the rest-frame of a moving particle. Their explicit expressions including the NNLO contribution
read
\begin{widetext}
\bse
\label{gyro}
\begin{align}
g^\mathrm{ADM}_{S}
&= 2 + \frac{1}{c^2} \bigglb(
\frac{19}{8}\nu\,\pp
+ \frac{3}{2}\nu\,\np^2
- \Big(6+2\nu\Big) \frac{1}{r} \biggrb)\nonumber\\
&+\dfrac{1}{c^4}
\Bigg\{-\dfrac{9}{8}\nu\Big(1-\dfrac{22}{9}\nu\Big)\p^4
        -\dfrac{3}{4}\nu\Big(1-\frac{9}{4}\nu\Big)\pp\np^2
        +\dfrac{15}{16}\nu^2\np^4\nonumber\\
        &\qquad\quad+\dfrac{1}{r}\bigg[ -\dfrac{157}{8}\nu\Big(1+\dfrac{39}{314}\nu\Big)\pp
                            - 16\nu\Big(1+\dfrac{45}{256}\nu\Big)\np^2 
                            + \frac{1}{r}\frac{21}{2}\Big(1+\nu\Big)\bigg]
\Bigg\},
\\
g^\mathrm{ADM}_{S^*}
&= \frac{3}{2} + \frac{1}{c^2} \bigglb(
\Big(-\frac{5}{8}+2\nu\Big)\pp
+ \frac{3}{4}\nu\,\np^2
- \Big(5+2\nu\Big) \frac{1}{r} \biggrb)\nonumber\\
&+\dfrac{1}{c^4}
\Bigg\{\dfrac{1}{16}\Big(7-37\nu+39\nu^2\Big)\p^4+\dfrac{9}{16}\nu(2\nu-1)\pp\np^2\nonumber\\
&\qquad\quad+\dfrac{1}{r}\bigg[\dfrac{1}{8}\Big(27-129\nu-\dfrac{39}{2}\nu^2\Big)\pp
-6\nu\Big(1+\dfrac{15}{32}\nu\Big)\np^2+\dfrac{1}{r}\left(\dfrac{75}{8}+\dfrac{41}{4}\nu\right)\bigg]
\Bigg\}.
\end{align}
\ese
\end{widetext}
The label ``ADM'' on the gyro-gravitomagnetic ratios~\eqref{gyro} is a reminder that, although
the LO values are coordinate independent,  both the NLO and NNLO contributions to these ratios actually 
depend on the definition of the phase-space variables $(\r,\p)$.
In the next Section we shall introduce the two, related, effective gyro-gravitomagnetic ratios 
that enter the effective EOB Hamiltonian, written in effective (or EOB) coordinates, according 
to the prescriptions of~\cite{Damour:2008qf}.

\section{Effective Hamiltonian and  effective gyro-gravitomagnetic ratios}
\label{sec:Heff}
Following Ref.~\cite{Damour:2008qf}, two operations have to be performed on the Hamiltonian
written in the center of mass  frame so to cast it in a form that can be resummed
in a way compatible to previous EOB work.
First of all, one needs to transform the (ADM) phase-space coordinates $(\x_a,\p_a,{\mathbf S}_a)$ 
by a canonical transformation compatible with the one used in previous EOB work.
Second, one needs  to compute the {\it effective} Hamiltonian corresponding to the
canonically transformed {\it real} Hamiltonian.
Following the same procedure adopted in~\cite{Damour:2008qf}, we start by performing the 
purely orbital canonical transformation which was found to be needed to go from 
the ADM coordinates used in the PN-expanded Hamiltonian to the coordinates used 
in the EOB dynamics.
Since in Ref.~\cite{Damour:2008qf} one was concerned only with the NLO spin-orbit 
interaction, it was enough to consider the 1PN-accurate transformation. 
In the present study, because one is working at NNLO in the spin-orbit interaction, 
one needs to take into account the complete 2PN-accurate canonical transformation 
introduced in~\cite{Buonanno:1998gg}.
The transformation changes the ADM phase-space variables $(\r,\p,\S,\Sstar)$ 
to $(\r',\p',\S,\Sstar)$ and it is explicitly given by Eqs.~(6.22)-(6.23) 
of~\cite{Buonanno:1998gg}. 
To our purpose, we actually need to use  the {\it inverse} relations 
$\r=\r(\r',\p')$ and  $\p=\p(\r',\p')$, so to replace $(\r,\p)$ 
with $(\r',\p')$ in Eq.~\eqref{eq:Hso}. 
The needed transformation is easily found by solving, by iteration, 
Eqs.~(6.22)-(6.23) of~\cite{Buonanno:1998gg}, and we explicitly quote 
it here for future convenience. It reads
\begin{widetext}
\begin{align}
\label{eq:canon_r}
r_i - r_i' &=\frac{1}{c^2}\Bigg[ -\left( 1+\dfrac{\nu}{2} \right)\dfrac{r'^i}{r'} + \dfrac{\nu}{2}\p'^2r_i'  + \nu (\r'\cdot \p')p'_i \Bigg]\nonumber \\
           & + \dfrac{1}{c^4}\Bigg\{ \Bigg[ \dfrac{1}{4r'^2}\left( -\nu^2 + 7\nu -1\right)   +\dfrac{3\nu}{4}\left(\dfrac{\nu}{2}-1\right)\dfrac{\p'^2}{r'}  
             - \dfrac{\nu}{8}\left( 1+\nu \right)\p'^4-\nu\left( 2+\dfrac{5}{8}\nu\right)\dfrac{(\r'\cdot\p')^2 }{r'^3}\Bigg]r'_i \nonumber\\
           & + \Bigg[\dfrac{\nu}{2} \left(-5+\dfrac{\nu}{2}\right) \dfrac{\r'\cdot\p'}{r'}+\dfrac{\nu}{2}\left(\nu-1\right)\p'^2 (\r'\cdot\p')\Bigg]p'_i  \Bigg\},
\end{align}
\begin{align}
\label{eq:canon_p}
p_i - p_i' &=  \frac{1}{c^2}\Bigg[-\left(1+\frac{\nu}{2}\right)\dfrac{\r'\cdot \p'}{r'^3}r'_i+\left(1+\dfrac{\nu}{2}\right)\dfrac{p_i'}{r'}-\dfrac{\nu}{2}\p'^2 p_i'\Bigg]\nonumber\\
           & + \frac{1}{c^4}\Bigg\{\Bigg[\dfrac{1}{r'^2}\left(\frac{5}{4}-\frac{3}{4}\nu+\frac{\nu^2}{2}\right) +\frac{\nu}{8} (1+3\nu)\p^4 
             -\dfrac{\nu}{4}\left(1+\dfrac{7}{2}\nu\right)\dfrac{\p^2}{r'}+\nu\left(1+\frac{\nu}{8}\right)\frac{(\r' \cdot \p')^2}{r'^3}\Bigg]p'_i \nonumber\\
           & +\Bigg[\left(-\frac{3}{2}+\frac{5}{2}\nu-\frac{3}{4}\nu^2\right)\dfrac{\r'\cdot \p'}{r'^4}
             + \dfrac{3}{4}\nu\left(\dfrac{\nu}{2}-1\right)\p^2 \dfrac{\r'\cdot\p'}{r'^3} + \dfrac{3}{8}\nu^2\dfrac{(\r'\cdot\p')^3}{r'^5}\Bigg]r'^i \Bigg\}.
\end{align}
\end{widetext}
As pointed out in~\cite{Buonanno:1998gg}, in the test-mass limit ($\nu\rightarrow 0$) one 
has $r'^i=\left[1+1/(2 c^2 r)\right]r^i$, 
which is the relation between Schwarzschild ($r'$) and isotropic ($r$) coordinates in a Schwarzschild spacetime\footnote{As a check
of the transformation~\eqref{eq:canon_r}-\eqref{eq:canon_p} one can explicitly verify that it preserves the orbital angular momemntum 
at 2PN order, i.e. $r'\times\p'=r\times\p+{\cal O}\left(\dfrac{1}{c^6}\right)$.}.
When this transformation is applied to to the spin-orbit Hamiltonian in ADM coordinates, Eq.~\eqref{eq:Hso}, 
one gets a transformed Hamiltonian of the form $\H'(\r',\p',\S,\Sstar)=\H'_{\rm o}(\r',\p',\S,\Sstar)+\H^{\prime \rm so}(\r',\p',\S,\Sstar)$,
with the NNLO spin-orbit contribution that explicitly reads
\begin{widetext}
\begin{align}
  \H^{\prime\rm so}_{\rm NNLO}(\r',\p',\S,\Sstar)= \dfrac{\nu}{r'^2}\Bigg\{
&(\bar{S}^*,n',p')\Bigg[\dfrac{\nu}{r'^2}\left(-8+\dfrac{\nu}{2}\right)\nonumber\\
&                      +\dfrac{1}{r'}\left[
                                          \left(-\dfrac{13}{4}\nu - \dfrac{3}{4}\nu^2\right)\p'^2
                                          +\left(\dfrac{43}{4}\nu-\dfrac{75}{16}\nu^2\right)(\n'\cdot\p')^2
                                     \right]
\nonumber\\
 &                     +\left(-\dfrac{3}{8}\nu+\dfrac{9}{16}\nu^2\right)\p'^4  
                       + \left(\dfrac{9}{4}\nu-\dfrac{3}{16}\nu^2\right)\p'^2(\n'\cdot\p')^2 
                       + \dfrac{135}{16}\nu^2(\n'\cdot\p')^2 \Bigg],\nonumber\\
&+(\bar{S}^*,n',p')
\Bigg[-\dfrac{1}{r'^2}\left(\dfrac{1}{2}+\dfrac{53}{8}\nu+\dfrac{5}{8}\nu^2\right)\nonumber\\
&                    +\dfrac{1}{r'}\left[
                                         \left(\dfrac{1}{4}-\dfrac{53}{16}\nu+\dfrac{3}{8}\nu^2\right)\p'^2
                                        +\left(\dfrac{5}{4}+\dfrac{121}{8}\nu-3\nu^2\right)(\n'\cdot\p')^2
                                   \right]
\nonumber\\
&                     +\left(\dfrac{7}{16}-\dfrac{3}{16}\nu+\dfrac{\nu^2}{4}\right)\p'^4 
                      + \left(\dfrac{57}{16}\nu-\dfrac{3}{4}\nu^2\right)\p'^2(\n'\cdot\p')^2  
                      + \dfrac{15}{2}\nu^2(\n'\cdot\p')^2\Bigg]
\Bigg\},
\end{align}
\end{widetext}
where we introduced the radial unit vector $\n'=\r'/|r'|$.

With this result in hands, we can further perform on it a secondary 
{\it purely spin-dependent}, canonical transformation that affects both the 
NLO and NNLO spin orbit terms. This transformation can be thought as a  gauge 
transformation related to the arbitrariness in choosing a spin-supplementary condition 
and in defining a local frame to measure the spin vectors. Such gauge condition 
can then be conveniently chosen so to simplify the spin-orbit Hamiltonian.
This procedure was pushed forward, at NLO accuracy in Ref.~\cite{Damour:2008qf}. 
In that case, the canonical transformation was defined by means of a 2PN-accurate 
generating function, that was chosen proportional to the spins and with two arbitrary 
($\nu$-dependent) dimensionless  coefficients $a(\nu)$ and $b(\nu)$. 
Using rescaled variables, the NLO generating function of~\cite{Damour:2008qf} reads 
\begin{align}
\label{eq:GSNLO}
\G_{\rm s 2PN}=\dfrac{1}{c^4}\nu \dfrac{(\n'\cdot \p')}{r'}\left(a(\nu)(\bar{S},n',p')+b(\nu)(\bar{S}^*,n',p')\right).
\end{align}
In Ref.~\cite{Damour:2008qf} the parameters $a(\nu)$ and $b(\nu)$ were selected so to remove the terms
proportional to $\p^2$ in the final (effective) Hamiltonian. Let us recall that, at linear order in the 
$\G_{\rm s 2PN}$, that was enough for the NLO case, the new Hamiltonian was computed as 
$\H^{\prime\prime \rm so}(y'')=\H^{\prime \rm so}(y'')-\{\H',\G_{\rm s2PN}\}(y'')$, were we address 
collectively with $y''=(\r'',\p'',\S'',{\bf S}^{\prime\prime *})$ the new phase space-variables.

We wish now to introduce a more general gauge transformation such to act also on the NNLO terms 
of the Hamiltonian. To do so, in addition to the NLO part $\G_{\rm s2PN}$ of the spin-dependent 
generating function mentioned above, one also needs to introduce  a NNLO contribution of the form
\begin{align}
\G_{\rm s3PN}=\dfrac{1}{c^6}\nu \Bigg\{&\dfrac{(\n'\cdot \p')}{r'}\left[\dfrac{\alpha(\nu)}{r'}+\beta(\nu)(\n'\cdot \p')^2+\gamma(\nu)\p'^2\right]\nonumber\\
                       &\times (\bar{S},n',p')\nonumber\\
                       +&\dfrac{(\n'\cdot \p')}{r'}\left[\dfrac{\delta(\nu)}{r'}+\zeta(\nu)(\n'\cdot \p')^2+\eta(\nu)\p'^2\right]\nonumber\\
                       &\times(\bar{S}^*,n',p')\Bigg\},
\end{align}
with six, arbitrary, $\nu$-dependent dimensionless coefficients. We shall then 
consider the effect of a spin-dependent generating function of the form 
$\G_s=\G_{\rm s 2PN}+\G_{\rm s3PN}$. Since $\G_s$ starts at 2PN order, it turns out 
that possible quadratic terms in the generating function are of order $c^{-8}$, 
i.e. at 4PN and thus are of higher order than the NNLO accuracy that we are 
currently considering in the spin-orbit Hamiltonian. The  consequence is that 
the purely spin-dependent gauge transformation at NNLO will involve {\it only} 
the contribution linear in $\G_s$. In other terms, we only need to consider 
the following transformation on the Hamiltonian
\be
\label{eq:Hpp}
\H^{\prime\prime }(y^{\prime\prime})=\H'(y^{\prime\prime})-\{\H',\G_s\}(y^{\prime\prime}).
\ee
Extracting from this equation the spin-dependent terms, we find that the
relevant terms in the new spin-orbit Hamiltonian up to NNLO are then given by
\begin{align}
\H^{\prime\prime{\rm so}}_{\rm LO}(\r^{\prime\prime},\p^{\prime\prime},\S^{\prime\prime},\S^{\prime\prime*})&=H^{\prime{\rm so}}_{\rm LO}(y^{\prime\prime}),\nonumber\\
\H^{\prime\prime{\rm so}}_{\rm NLO}(\r^{\prime\prime},\p^{\prime\prime},\S^{\prime\prime},\S^{\prime\prime*})&=
H^{\prime{\rm so}}_{\rm NLO}(y^{\prime\prime})-\{H^{\prime}_{\rm oN},\G_{\rm s2PN}\}(y^{\prime\prime}),\nonumber\\
\label{hpp:NNLO}
\H^{\prime\prime{\rm so}}_{\rm NNLO}(\r^{\prime\prime},\p^{\prime\prime},\S^{\prime\prime},\S^{\prime\prime*})&=\H'^{\rm so}_{\rm NNLO}(y^{\prime\prime})\nonumber\\
&-\Big[\{\H'_{\rm oN},\G_{\rm s3PN}\}\nonumber\\
&+\{\H'_{\rm o1PN},\G_{\rm s2PN}\}\nonumber\\
&+\{H_{\rm LO}^{\prime\rm so},\G_{\rm s2PN}\}\Big](y^{\prime\prime} ).
\end{align}
Note that the single prime in these equations explicitly addresses the various contribution 
to the spin-orbit Hamiltonian as  computed after the purely orbital canonical transformation 
mentioned above (note however that only the functional form of  $\H_{\rm o1PN}'$ is modified by 
the action of the orbital canonical transformation). 

Further simplifications 
occur in the third Poisson bracket of Eq.~\eqref{hpp:NNLO}.
First of all, since we are interested in computing only the contribution to the
spin-orbit interaction, the terms quadratic in spins are neglected. 
In addition, from the basic relation $\{S_i,S_j\}=\epsilon_{ijk}S_k$ one can show 
by a straightforward calculation that $\{H_{\rm LO}^{\rm so\prime},\G_{\rm s2PN}\}=0$
(always at linear order in the spin). Consequently, the effect of the purely 
spin-dependent canonical transformation is fully taken into account by the two 
Poisson brackets involving the generating functions $\G_{\rm s2PN}$ and $\G_{\rm s3PN}$, 
and the purely orbital contributions to the Hamiltonian, $\H'_{\rm oN}$ and $\H'_{\rm o1PN}$.

For simplicity of notation, we shall omit hereafter the double primes from the
transformed Hamiltonian. We now need to connect the real Hamiltonian 
$H$ to the effective one $H_{\rm eff}$, which is more closely linked to the 
description of the EOB quasigeodesic dynamics.
The relation between the two Hamiltonians is given by~\cite{Buonanno:1998gg}
\be
\dfrac{H_{\rm eff}}{\mu c^2}\equiv \dfrac{H^2-m_1^2 c^4 - m_2^2 c^4}{2m_1m_2 c^4}
\ee
where the real Hamiltonian $H$ contains the rest-mass contributions $Mc^2$.
In terms of the nonrelativistic Hamiltonian $\hat{H}^{\rm NR}$, this equation 
is equivalent to
\be
\dfrac{\hat{H}_{\rm eff}}{c^2}=1 + \dfrac{\H^{\rm NR}}{c^2}+\dfrac{\nu}{2}\dfrac{(\H^{\rm NR})^2}{c^4},
\ee
where it is explicitly
\begin{align}
\H^{\rm NR}=\left(\H_{\rm oN} + \dfrac{\H_{\rm o1PN}}{c^2}+\dfrac{\H_{\rm o2PN}}{c^4}+\dfrac{\H_{\rm o3PN}}{c^6}\right) \nonumber\\
          +\left(\dfrac{\H_{\rm LO}^{\rm so}}{c^2}+\dfrac{\H_{\rm NLO}^{\rm so}}{c^4}+\dfrac{\H_{\rm NNLO}^{\rm so}}{c^6}\right).
\end{align}
By expanding in powers of $1/c^2$ up to 3PN fractional accuracy (and in powers of the spin) the exact 
effective Hamiltonian, one easily finds that the spin-orbit part of the effective Hamiltonian $\H_{\rm eff}$ 
(i.e., the part which is linear-in-spin) reads
\begin{align}
\H_{\rm eff}^{\rm so}&=\dfrac{1}{c^2}\H_{\rm LO}^{\rm so} + \dfrac{1}{c^4}\left(\H_{\rm NLO}^{\rm so} + \nu\H_{\rm oN}\H_{\rm LO}^{\rm so}\right)\nonumber\\
                  &+\dfrac{1}{c^6}\Big[\H^{\rm so}_{\rm NNLO}+\nu\left(\H_{\rm oN}\H_{\rm NLO}^{\rm so}+\H_{\rm o1PN}H_{\rm LO}^{\rm so}\right)\Big].
\end{align}

Combining this result with the effect of the generating function discussed above, we get the transformed
spin-orbit part of the effective Hamiltonian in the form as
\be
\label{eq:Hsoeff}
\H^{\rm so}_{\rm eff}=\dfrac{\nu}{c^2 r^2}\left(g_S^{\rm eff}(\bar{S},n,p) + g_{S^*}^{\rm eff}(\bar{S}^*,n,p)\right).
\ee
The effective gyro-gravitomagnetic ratios $g_S^{\rm eff}$ and $g_{S^*}^{\rm eff}$ differ from the ADM ones introduced 
above because of the effect of the (orbital+spin) canonical transformation and because of the transformation 
from $H$ to $H_{\rm eff}$. They have the structure
\begin{align}
g_S^{\rm eff}     = 2 & + \dfrac{1}{c^2}g_S^{\rm eff_{\rm NLO}}(a) + \dfrac{1}{c^4}g_S^{\rm eff_{\rm NNLO}}(a;\alpha,\beta,\gamma) \\
g_{S^*}^{\rm eff}  = \dfrac{3}{2} &+ \dfrac{1}{c^2}g_S^{\rm eff_{\rm NLO}}(b) + \dfrac{1}{c^4}g_{S^*}^{\rm eff_{\rm NNLO}}(b;\delta,\zeta,\eta), 
\end{align}
where we made it apparent the dependence on the ($\nu$-dependent) NLO and NNLO gauge parameters.
Including the new NNLO terms, they read
\begin{widetext}
\begin{align}
 \label{eq:gSeff_full}
  g_S^{\rm eff}=2&+\dfrac{1}{c^2}\Bigg[\left(\dfrac{3}{8}\nu+a\right)\p^2-\left(\dfrac{9}{2}\nu+3 a\right)(\n\cdot\p)^2\Bigg)
                -\dfrac{1}{r}(\nu+a)\Bigg]\nonumber\\
               &+\dfrac{1}{c^4}\Bigg[-\dfrac{1}{r^2}\left(9\nu + \dfrac{3}{2}\nu^2 + a + \alpha \right)
                   \nonumber\\
                  &+\dfrac{1}{r}\left[(\n\cdot\p)^2 \left(\dfrac{35}{4}\nu - \dfrac{3}{16}\nu^2 + 6a - 4\alpha-3\beta-2\gamma\right)
                   +\p^2\left(-\dfrac{17}{4}\nu + \dfrac{11}{8}\nu^2 - \dfrac{3a}{2}+\alpha-\gamma\right)\right]\nonumber\\
                   &+ \left(\dfrac{9}{4}\nu - \dfrac{39}{16}\nu^2 + \dfrac{3a}{2} +3\beta-3\gamma \right)   \p^2 (\n\cdot \p)^2
                    + \left(\dfrac{135}{16}\nu^2-5\beta\right)                                              (\n\cdot \p)^4   \nonumber \\
                   &+\left(-\dfrac{5}{8}\nu-\dfrac{a}{2} + \gamma\right)                                    \p^4             
                    \Bigg],\\  
 g_{S^*}^{\rm eff}  =\dfrac{3}{2}&+\dfrac{1}{c^2}\Bigg[\left(-\dfrac{5}{8}+\dfrac{1}{2}\nu+b\right)\p^2 
                 -\left(\dfrac{15}{4}\nu+3b\right)(\n\cdot\p)^2-\dfrac{1}{r}\left(\dfrac{1}{2}+\dfrac{5}{4}\nu+b\right)\Bigg]\nonumber\\ 
    &+\dfrac{1}{c^4}\Bigg[-\dfrac{1}{r^2}\left(\dfrac{1}{2} + \dfrac{55}{8}\nu + \dfrac{13}{8}\nu^2 + b+ \delta\right)
                     \nonumber\\
                   &+\frac{1}{r}\Bigg[(\n\cdot \p)^2\left(\dfrac{5}{4}+\dfrac{109}{8}\nu + \dfrac{3}{4}\nu^2+6 b-4\delta-3\zeta-2\eta\right)
                    +\p^2\left(\dfrac{1}{4}-\dfrac{59}{16}\nu + \dfrac{3}{2}\nu^2 - \dfrac{3b}{2}+ \delta-\eta\right)\Bigg]\nonumber\\
                   & +\left(\dfrac{57}{16}\nu - \dfrac{21}{8}\nu^2 + \dfrac{3b}{2} + 3\zeta - 3\eta\right)     \p^2(\n\cdot\p)^2
                    +\left(\dfrac{15}{2}\nu^2                       - 5\zeta\right)                            (\n\cdot \p)^4\nonumber\\
                   &+\left(\dfrac{7}{16}-\dfrac{11}{16}\nu-\dfrac{\nu^2}{16}-\dfrac{b}{2} + \eta \right) \p^4
                    \Bigg].
\label{eq:gSstareff_full}
\end{align}
\end{widetext}
This is the central result of the paper. The NNLO contribution to the gyro-gravitomagnetic 
ratios computed here is the crucial, new, information that it is needed to improve
to the next PN order the spin-dependent EOB Hamiltonian (either in the version of
Ref.~\cite{Damour:2008qf} or~\cite{Barausse:2009xi}).
Let us recall in this respect that in the EOB approach of~\cite{Damour:2008qf}
the relative dynamics can be  equivalently represented by the dynamics of a 
spinning effective particle with effective spin $\bsigma$ 
moving onto a $\nu$-deformed Kerr-type metric. The gyro-gravitomagnetic ratios 
enter the definition of the test-spin vector $\bsigma$ as
\be
\label{def:sigma}
\bsigma=\dfrac{1}{2}\left(g_S^{\rm eff}-2\right){\bf S} + \dfrac{1}{2}\left(g_{S^*}^{\rm eff}-2\right){\bf S}^*,
\ee
that can then be inserted in Eqs.~(4.16) of Ref.~\cite{Damour:2008qf} to get the spin-orbit interaction
additional to the leading Kerr-metric part. Together with Eqs.~(4.17), (4.18) and (4.19) of 
Ref.~\cite{Damour:2008qf}  this defines the real EOB-improved, resummed Hamiltonian for 
spinning binaries at NNLO in the spin-orbit interaction.

\section{Limits, checks and gauge fixing}
\label{sec:fixings}

\subsection{The extreme-mass-ratio limit}
\label{sbsc:EMRL}
The effective spin-orbit Hamiltonian~\eqref{eq:Hsoeff} is naturally connected 
to the test-mass ($\nu\to 0$)
Hamiltonian explicitly obtained\footnote{Note in passing that the simple procedure 
described in Ref.~\cite{Damour:2007nc} 
to obtain the spin-orbit Hamiltonian is totally general and can be applied, in particular, to 
the test-mass case.} in~\cite{Barausse:2009aa}. 
To show this in a concrete case, let us consider the spin-orbit Hamiltonian of a spinning test-particle
on Schwarzschild spacetime written explicitly using isotropic coordinates, as given by Eq.~(5.12) of 
Ref.~\cite{Barausse:2009aa}. By considering the Schwarzschild 
metric written as
\be
ds^2 = -f(r) dt^2 + h(r)(dx^2 + dy^2 + dz^2),
\ee
where $r$ labels here the isotropic radius\footnote{Note that we use the same notation for the isotropic radius on
Schwarzschild spacetime and the ADM radial coordinates. There is no ambiguity here since for the Schwarzschild 
spacetime ADM coordinates do actually coincide with isotropic coordinates}, $r^2=x^2+y^2+z^2$, (that is
meant to be expressed in rescaled units, where now $M\simeq m_1$ is the background mass and $\mu\simeq m_2$ is the
test-particle mass), with 
\be
h=\left(1+\dfrac{1}{2c^2 r}\right)^4,
\ee
and using rescaled variables (and making explicit the speed of light) Eq.~(5.12) of 
Ref.~\cite{Barausse:2009aa} can be written as
\begin{equation}
\label{eq:so_test}
\hat{H}^{\rm so}_{\rm ISO}=\dfrac{\nu}{c^2 r^2}g_0^{\rm ISO}\left(n,p,\bar{S}^*_0\right).
\end{equation}
In this equation, $\S^*_0$ is the (rescaled) spin of the test-mass and we have introduced
the test-mass gyro-gravitomagnetic ratio in isotropic coordinates $g_0^{\rm ISO}$, that is 
known in closed form~\cite{Barausse:2009aa} and reads
\be
g_0^{\rm ISO}=\dfrac{h^{-3/2}}{\sqrt{Q}\left(1+\sqrt{Q}\right)}\left[1-\dfrac{1}{2c^2 r} + \left(2-\dfrac{1}{2c^2 r}\right)\sqrt{Q}\right],
\ee
where 
\be
Q= 1 + \dfrac{1}{c^2}\dfrac{\p^2}{h}.
\ee
By transforming the Hamiltonian~\eqref{eq:so_test} from isotropic to Schwarzschild coordinates
using the $\nu\to 0$ limit of the (purely orbital) canonical transformation given by
Eqs.~\eqref{eq:canon_r}-\eqref{eq:canon_p}, expanding in powers of $1/c^2$, (and dropping
again the primes for simplicity) one obtains
\be
\hat{H}^{\rm so}_{\rm Schw}=\dfrac{\nu}{c^2 r^2}g_0^{\rm Schw}\left(n,p,\bar{S}^*\right).
\ee
with
\begin{align}
\label{eq:g0}
g_0^{\rm Schw}&=\dfrac{3}{2}-\dfrac{1}{c^2}\left(\dfrac{1}{2r}+\dfrac{5}{8}\p^2 \right)\nonumber\\
            &+\dfrac{1}{c^4}\left[-\dfrac{1}{2r^2}+\dfrac{1}{r}\left( \dfrac{5}{4}(\n\cdot\p)^2 + \dfrac{1}{4}\p^2 \right)+\dfrac{7}{16}\p^4\right].
\end{align}
In the $\nu\to 0$ (Schwarzschild) limit, one has
$\lim_{\nu\to 0}(H-{\rm const.})/\mu=\lim_{\nu\to 0}\H_{\rm eff}$ 
(when dropping inessential constants), $\Sstar=\Sstar_0$ and $\S= 0$. 
One then finds that the result~\eqref{eq:g0} agrees in the $\nu\to 0$ limit 
with Eq.~\eqref{eq:gSstareff_full} when the gauge parameters 
$(b,\delta,\zeta,\eta)$ are simply zero.

In addition, in the $\nu\to 0$ limit where the background is a Kerr black hole, 
i.e. $\S\neq 0$, Eq.~\eqref{eq:gSeff_full} consistently exhibits that both the 
NLO and NNLO contributions become pure gauge, that can just be set to zero 
by demanding $(a,\alpha,\beta,\gamma)$ to vanish.

\subsection{Circular equatorial orbits}
\label{sec:circular}
Let us consider now the situation where both individual spins are parallel 
(or antiparallel) to the (rescaled) orbital angular momentum vector 
$\bl=r\n\times\p$. [Note that in this Section the quantity $r$ denotes the 
EOB radial coordinate (further modified by spin-dependent gauge terms, see below)].
In this case, circular orbits exists (but in the general case, when the 
spin vectors are not aligned with $\bl$, there are no circular orbits).
One can then consistently set everywhere the radial momentum to zero, 
$p_r\equiv \n\cdot\p=0$ and express the total (orbital plus spin-orbit part) 
real, PN-expanded and canonically transformed Hamiltonian, 
$H(y'')\equiv H''_{\rm o}(y'')  + H^{\prime\prime \rm so}(y'')$ (dropping hereafter the
primes for simplicity) as a function of $r$, $\ell$ (using the link $\p^2=\ell^2/r^2$, 
where $\ell\equiv |\bl|$) and of the two scalars $\hat{a}$ and $\hat{a}^*$ 
measuring the projection of the basic spin combinations ${\bf S}$ and ${\bf S}^*$ along 
the direction of the orbital angular momentum $\bl$. 
Following the same notation of~\cite{Damour:2008qf}, 
we introduce here the dimensionless spin variables corresponding to ${\bf S}$ and ${\bf S}^*$ 
\be
\hat{\bf a}\equiv \dfrac{c {\bf S}}{GM^2},\qquad \hat{\bf a}^*\equiv \dfrac{c{\bf S}^*}{GM^2},
\ee
and we define the projections as
\be
\hat{\bf a}\cdot\bl =\ah\ell,\qquad
\hat{\bf a}^*\cdot\bl=\ahstar\ell,
\ee
with the scalars $\ah$ and $\ahstar$ positive or negative depending on whether
say $\hat{\bf a}$ is parallel or antiparallel to $\bl$. The sequence of 
circular (equatorial) orbits\footnote{To avoid confusion, let us stress that we are here considering the circular 
orbits of the PN-expanded real Hamiltonian and {\it not} the circular orbits of the EOB-resummed real Hamiltonian, 
as done in Sec.~V of Ref.~\cite{Damour:2008qf}. This analysis is postponed to future work.} 
is then determined by the constraint 
\begin{equation}
\label{eq:dH0}
\de H(r,\ell,\ah,\ahstar)/\de r=0,
\end{equation}
(or equivalently by $\de H_{\rm eff}/\de r=0$).
To start with, let us consider first the link between the nonrelativistic 
energy (per unit mass $\mu$) and the orbital angular momentum along circular orbits. 
The relevance of this quantity in the nonspinning case, say 
$E_{\rm circ}(\ell)\equiv H_{\rm o}^{\rm NR}(\ell)/\mu$,  was  pointed out in 
Ref.~\cite{Damour:1999cr}, since it provides a completely gauge-invariant 
characterization of the dynamics of circular orbits. 
When the black holes are spinning, the same property of gauge-invariance is 
maintained when the spins are parallel (or antiparallel) to the orbital angular 
momentum, so that it is meaningful to explicitly compute 
$E_{\rm circ}(\ell,\ah,\ahstar)\equiv  H_{\rm o}^{\rm NR}(\ell)/\mu + H_{\rm so}^{\rm NR}(\ell,\ah,\ahstar)/\mu$ 
in this case. 
Since it is a gauge-invariant quantity, the result is independent
of the canonical transformations that we have performed on the two-body Hamiltonian
in ADM coordinates, so that it gives a reliable check of the procedure we followed.
As a first operation, we need to solve, iteratively, the constraint~\eqref{eq:dH0} so 
to obtain the EOB coordinate radius $r$ in function of $(\ell,\ah,\ahstar)$. This
function (that is not invariant and depends explicitly on the gauge parameters) reads
(putting back the explicit double primes on $r$ as a reminder that this is the
EOB radial coordinate)
\begin{widetext}
\begin{align}
\label{eq:r_ell}
r^{\prime\prime}(\ell,\ah,\ahstar)&=\ell^2\Bigg\{1+\dfrac{1}{c^2}\left[-\dfrac{3}{\ell^2}+\dfrac{1}{c}\dfrac{1}{\ell^3}\left(6\,\ah + \dfrac{9}{2}\,\ahstar\right)\right]\nonumber\\
&+\dfrac{1}{c^4}\Bigg[(-9+3\nu)\dfrac{1}{\ell^4} +\dfrac{1}{c}\dfrac{1}{\ell^5}\Bigg(  
    \ah\left(33-\dfrac{17}{8}\nu+a(\nu)\right) + \ahstar\left(\dfrac{157}{8}-\dfrac{5}{2}\nu+b(\nu) \right) \Bigg)  \Bigg]\nonumber\\
&                +\dfrac{1}{c^6}\Bigg[\left(-54 + \dfrac{257}{3}\nu - \dfrac{41}{16}\pi^2\nu\right)\dfrac{1}{\ell^6}
                 + \dfrac{1}{c}\dfrac{1}{\ell^7}\Bigg(    \ah\left( \dfrac{1197}{4}  - \dfrac{1973}{16}\nu + \dfrac{3}{4}\nu^2   + 6a(\nu) + \alpha(\nu) + \gamma(\nu) \right) \nonumber\\ 
                     &    +\ahstar\left( \dfrac{2777}{16} - \dfrac{1633}{16}\nu + \dfrac{7}{16}\nu^2  + 6b(\nu) + \delta(\nu) +   \eta(\nu) \right)  
\Bigg)    \Bigg]\Bigg\}.
\end{align}
\end{widetext}
The function $E_{\rm circ}(\ell,\ah,\ahstar)$ is obtained by inserting this 
relation in the expression of $H(r,\ell,\ah,\ahstar)$, and it reads
\begin{widetext}
\begin{align}
\label{eq:Ecirc}
E_{\rm circ}(\ell,\ah,\ahstar)=&-\dfrac{1}{2\ell^2}\Bigg\{1+ \dfrac{1}{c^2}\left(\dfrac{1}{4}(9+\nu)\dfrac{1}{\ell^2}-\dfrac{1}{c}\dfrac{1}{\ell^3}\left(4\ah+3\ahstar\right)
                                \right) \nonumber\\
                              &+\dfrac{1}{c^4}\Bigg[\dfrac{1}{8}\left(81-7\nu+\nu^2\right)\dfrac{1}{\ell^4} 
                                 -\dfrac{1}{c}\dfrac{1}{\ell^5}\left(\left(36+\dfrac{3}{4}\nu\right)\ah  +\dfrac{99}{4}\ahstar \right)\Bigg]\nonumber\\
          &+\dfrac{1}{c^6}\Bigg[ \dfrac{2}{\ell^6}o_1(\nu)
          +\dfrac{1}{c}\dfrac{1}{\ell^7}\left(\ah    \left(           -324  + 54\nu        -\dfrac{5}{8}\nu^2 \right) 
                  + \ahstar\left(-\dfrac{1701}{8} + \dfrac{195}{4}\nu               \right)   
\right) \Bigg]\Bigg\}
\end{align}
\end{widetext}
where we defined
\begin{equation}
2 o_1(\nu) = \dfrac{3861}{64}-\dfrac{8833}{192}\nu+\dfrac{41}{32}\pi^2\nu-\dfrac{5}{32}\nu^2+\dfrac{5}{64}\nu^3,
\end{equation}
for the 3PN-accurate orbital part, with a slight abuse of the notation of Ref.~\cite{Damour:1999cr}.
Note that, as it should, Eq.~\eqref{eq:Ecirc} is totally independent of the eight gauge parameters. 
We have further verified that that the same result~\eqref{eq:Ecirc} is obtained starting from the 
PN-expanded Hamiltonian written in ADM coordinates and in the center of mass  frame, 
Eqs.~\eqref{eq:Hso}-\eqref{gyro}.

As a last remark, let us note that, as it was the case at NLO~\cite{Damour:2008qf}, 
the effective gyro-gravitomagnetic ratios for circular orbits are gauge independent also at NNLO. 
To see this explicitly, one just imposes in Eqs.~\eqref{eq:gSeff_full}-\eqref{eq:gSstareff_full} 
the condition $(\n\cdot\p)=0$ and the (approximate) link
\be
\p^2 = \dfrac{1}{r}+\dfrac{1}{c^2}\dfrac{3}{r^2}+{\cal O}(\hat{a},\hat{a}^*),
\ee
that is obtained by inverting Eq.~\eqref{eq:r_ell} at 1PN accuracy and neglecting the linear-in-spin
terms (that would give quadratic-in-spin contributions). At NNLO, one obtains
\begin{align}
\label{eq:circ_one}
g_{\rm S_{\rm circ}}^{\rm eff}  = 2           &-\dfrac{1}{c^2}\dfrac{5}{8}\nu\dfrac{1}{r}
                                         -\dfrac{1}{c^4}\left(\dfrac{51}{4}\nu + \dfrac{1}{8}\nu^2\right)\dfrac{1}{r^2},\\
\label{eq:circ_two}
g_{\rm S^*_{\rm circ}}^{\rm eff} =\dfrac{3}{2} &-\dfrac{1}{c^2}\left(\dfrac{9}{8}+\dfrac{3}{4}\nu\right)\dfrac{1}{r}\nonumber\\
                                         &-\dfrac{1}{c^4}\left(\dfrac{27}{16}+\dfrac{39}{4}\nu+\dfrac{3}{16}\nu^2\right)\dfrac{1}{r^2}.
\end{align} 
These equations indicate that the inclusion of NNLO spin-orbit coupling has the effect 
of {\it reducing} the magnitude of the  gyro-gravitomagnetic ratios. The NNLO
and NLO spin-orbit contributions act then in the same direction, so to reduce the repulsive
effect of the LO spin-orbit coupling which is, by itself, responsible for allowing the
binary system to orbit on very close, and very bound, orbits (see also 
Ref.~\cite{Damour:2001tu} and the discussion in Sec.~VI of~\cite{Damour:2008qf}).
We postpone to future work a detailed quantitative analysis of the properties of the 
binding energy entailed by Eqs.~\eqref{eq:circ_one}-\eqref{eq:circ_two}.

\subsection{Gauge fixing}
\label{sec:gauge-fixing}
We can finally exploit the flexibility introduced by the spin-dependent gauge 
transformation so to considerably simplify the expression of the effective 
gyro-gravitomagnetic ratios, Eqs.~\eqref{eq:gSeff_full}-\eqref{eq:gSstareff_full}.
This is helpful in the study of the dynamics of a binary system with generically 
oriented spins. Reference~\cite{Damour:2008qf} found it convenient to fix the 
NLO gauge parameters $(a(\nu),b(\nu))$ to
\be
\label{eq:NLO_gauge}
a(\nu)=-\dfrac{3}{8}\nu,\qquad b(\nu)=\dfrac{5}{8}-\dfrac{\nu}{2}
\ee
so to suppress the dependence on $\p^2$ at NLO. One can follow the same route
at NNLO, i.e., by choosing the six gauge parameters so to suppress the terms 
proportional to $\p^2$, $\p^4$ and $\p^2(\n\cdot\p)^2$. In this way the spin-orbit
Hamiltonian is expressed in a way that the circular (gauge-invariant) part is immediately
recognizable. With $(a,b)$ fixed as per Eq.~\eqref{eq:NLO_gauge}, one easily sees that 
the aforementioned NNLO terms are removed by the following choices of the 
NNLO gauge parameters
\begin{eqnarray}
\alpha(\nu)&=&\dfrac{11}{8}\nu\left(3-\nu\right),\\
\nonumber\\
\beta(\nu) &=&\dfrac{1}{16}\nu\left(13\nu-2\right),\\
\nonumber\\
\gamma(\nu)&=&\dfrac{7}{16}\nu,\\
\nonumber\\
\delta(\nu)&=&\dfrac{1}{16}(9+54\nu-23\nu^2),\\
\nonumber\\
\eta(\nu)  &=&\dfrac{1}{16}\left(-2+7\nu + \nu^2\right),\\
\nonumber\\
\zeta(\nu) &=&\dfrac{1}{16}\left(-7 - 8\nu + 15\nu^2\right).
\end{eqnarray}
The effective gyro-gravitomagnetic ratios are then simplified to
\begin{widetext}
\begin{align}
  g_S^{\rm eff}   = 2+&\dfrac{1}{c^2}\left\{-\dfrac{1}{r}\dfrac{5}{8}\nu-\dfrac{27}{8}\nu(\n\cdot\p)^2\right\}\nonumber\\
                  +&\dfrac{1}{c^4}\left\{-\dfrac{1}{r^2}\left(\dfrac{51}{4}\nu+\dfrac{\nu^2}{8}\right)
                       +\dfrac{1}{r}\left(-\dfrac{21}{2}\nu +\dfrac{23}{8}\nu^2\right)(\n\cdot\p)^2
                       +\dfrac{5}{8}\nu\left(1+7\nu\right)(\n\cdot\p)^4\right\},\\
g_{S^*}^{\rm eff}  = \dfrac{3}{2}+&\dfrac{1}{c^2}\left\{-\dfrac{1}{r}\left(\dfrac{9}{8}+\dfrac{3}{4}\nu\right)-\left(\dfrac{9}{4}\nu+\dfrac{15}{8}\right)(\n\cdot\p)^2\right\}\nonumber\\
                  +&\dfrac{1}{c^4}\left\{-\dfrac{1}{r^2}\left(\dfrac{27}{16}+\dfrac{39}{4}\nu + \dfrac{3}{16}\nu^2\right)
                       +\dfrac{1}{r}\left(\dfrac{69}{16}-\dfrac{9}{4}\nu+\dfrac{57}{16}\nu^2\right)(\n\cdot\p)^2
                       +\left(\dfrac{35}{16}+\dfrac{5}{2}\nu+\dfrac{45}{16}\nu^2\right)(\n\cdot\p)^4\right\}.
\end{align}
\end{widetext}
This result extends the information of Eqs.~(3.15a) and (3.15b) of Ref.~\cite{Damour:2008qf} at NNLO accuracy. 
The circular-orbit result mentioned above is immediately recovered ``at sight'' by imposing $(\n\cdot\p)=0$.
With this result in hand, one can proceed similarly to Sec.~IV of Ref.~\cite{Damour:2008qf} (as outlined above) 
to introduce the spin-dependent EOB-resummed real Hamiltonian including NNLO spin-orbit couplings.

\section{Conclusions}
\label{sec:conclusion}
Building on the recently-computed next-to-next-to-leading order PN-expanded spin-orbit Hamiltonian for two 
spinning compact objects~\cite{Hartung:2011te}, we computed the effective gyro-gravitomagnetic ratios entering
the EOB Hamiltonian at next-to-next to-leading order in the spin-orbit interaction. 
This result is obtained by a straightforward extension of the procedure followed in~\cite{Damour:2008qf} to derive 
the NLO spin-orbit EOB Hamiltonian. We discussed in detail the test-particle limit and the case of equatorial circular
orbits, when the spins are parallel or antiparallel to the orbital angular momentum.
In this case, one finds that the NNLO spin-orbit terms moderate the effect of 
the spin-orbit coupling (as the NLO terms was already doing~\cite{Damour:2008qf}).

Finally, while this paper was under review process, Ref.~\cite{Barausse:2011ys} appeared in the archives as
a preprint: that study uses the Lie method to obtain effective gyro-gravitomagnetic coefficients that are 
physically equivalent to the ones presented here. In addition, it also works out two classes of EOB 
Hamiltonians that are different from the one considered here.

\acknowledgements
I am grateful to Thibault Damour for suggesting this project 
and for maieutic discussions during its developement.

\end{document}